\documentclass[runningheads]{llncs}

\usepackage{amssymb}
\usepackage{graphicx}
\usepackage{psfrag, epsfig,amssymb}
\def\EE{\mathbb{E}}
\def\PP{\mathbb{P}}
\def\NN{\mathbb{N}}

\def\ind{{\rm 1\hspace{-0.90ex}1}}

\def\o{{\O}}

\def\=d{\stackrel{d}{=}}

\def\tP{\tilde{P}}

\def\th{\tilde{h}}

\def\gen{{\rm{gen}}}

\usepackage{url}
\urldef{\mailsa}\path|{alfred.hofmann,ursula.barth,ingrid.beyer,natalie.brecht,|
\urldef{\mailsb}\path|christine.guenther,frank.holzwarth,piamaria.karbach,|
\urldef{\mailsc}\path|anna.kramer,erika.siebert-cole,lncs}@springer.com|
\newcommand{\keywords}[1]{\par\addvspace\baselineskip
\noindent\keywordname\enspace\ignorespaces#1}

\begin{document}

\mainmatter  

\title{Marketing in a random network}

\titlerunning{Marketing in a random network}

%
%
\author{Hamed Amini
\and Moez Draief \and Marc Lelarge}

\authorrunning{Marketing in Random Networks}
\institute{INRIA, ENS Paris and Imperial College London}

%
\maketitle

\begin{abstract}
Viral marketing takes advantage of preexisting social networks among
customers to achieve large changes in behaviour.
Models of influence spread have been studied in a
number of domains, including the effect of ``word of mouth'' in the
promotion of new products or the diffusion of technologies.
A social network can be represented by a graph where the nodes are
individuals and the edges indicate a form of social relationship.
The flow of influence through this network can be thought of as an
increasing process of active nodes: as individuals become aware of new
technologies, they have the potential to pass them on to their
neighbours.
The goal of marketing is to trigger a large cascade of adoptions.
In this paper, we develop a mathematical model that allows to analyze
the dynamics of the cascading sequence of nodes switching to the new
technology. To this end we describe a continuous-time and a discrete-time models and analyse the proportion of nodes that adopt the new technology over time.
\end{abstract}

\noindent
\section{Introduction}
With consumers showing increasing resistance to traditional forms of
advertising, marketers have turned to alternate strategies like {\em viral
marketing}. Viral marketing exploits existing social networks by encouraging
customers to share product information with their friends.
Social networks are graphs in which nodes represent individuals and
edges represent relations between them.
To illustrate viral marketing, consider a company
that wishes to promote its new instant messenger (IM) system \cite{mmb}. A promising way
would be through popular social network such as Myspace: by convincing
several persons to adopt the new IM system, the company can obtain an
effective marketing campaign and diffuse the new system over the network.

If we assume that ``convincing'' a person to ``spread'' the new technology
costs money, then a natural problem is to detect the influential
members of the network who can trigger a cascade of influence in the
most effective way \cite{dr01}, \cite{kl07}.
In this work, we consider a slightly different problem: the marketer
has no knowledge of the social network. Hence he will not be
able to detect the most influential individuals and his only solution
is to ``convince'' a fraction of the total population. However, the
marketer can still use the structure of the underlying network by
targeting the neighbours of the adopters.
There are a number of incentive programs around this idea: each time an individual chooses the new technology,
he is given the opportunity to send e-mail to friends with a special
offer; if the friend goes on to buy it, each of the individuals
receives a small cash bonus.

In this paper, we develop a mathematical model that allows to analyze
the dynamics of the cascading sequence of nodes switching to the new
technology. To this end we describe a continuous-time and a discrete-time models and analyze the proportion of nodes that adopt the new technology over time. In the continuous setting we derive a general bound for the proportion of new adopters in terms global graph properties, namely the spectral radius and the minimum degree. In the discrete setting we show that the proportion of new adapters is the solution of a fixed point equation. To this end we examine the case of regular trees, and prove that our approach carries over to random regular graphs. We extend our model to the general threshold model \cite{kl07} and to sparse random graphs.
We conclude by presenting a framework that enables the control of the marketing policy and discuss other possible applications.

\section{Model}\label{Model}

We consider a set of $n$ agents represented by an undirected graph structure
 $G = (V, E)$ accounting for their interaction. For
$i,j\in V$, we write $i\sim j$ if $(i,j)\in E$ and we say that agents
$i$ and $j$ are neighbours. As in  \cite{Watts}, we consider binary models where each agent may choose between two possible strategies that we denote by $A$ and $B$. Let us  introduce a game-theoretic diffusion model proposed by Morris \cite{mor}:  Whenever two neighbours in the graph opt for strategy $A$ they receive a payoff $q_A$, if they both choose $B$ they receive a payoff $q_B$, and they receive nothing if they choose opposite strategies. The payoff of an agent corresponds to the sum of its payoffs with each of its neighbours.

Initially all nodes play $A$ except for a small number of nodes that are forced to adopt strategy $B$. The nodes that started with strategy $A$ will subsequently apply
best-response updates. More precisely, these nodes will be
repeatedly applying the following rule: switch to $B$ if enough of
your neighbours have already adopted $B$. There can be a cascading
sequence of nodes switching to $B$ such that a network-wide
equilibrium is reached in the limit.
This equilibrium may involve uniformity with all nodes adopting $B$ or
it may involve coexistence, with the nodes partitioned into a set
adopting $B$ and a set sticking to $A$.

The state of agent $i$ is represented by
$X_i$; $X_i=0$ if player $i$ plays strategy $A$ and
$X_i=1$ otherwise.
Hence $\sum_{j\sim i}X_j$ is the number of neighbours of $i$ playing
strategy $B$ and $\sum_{j\sim i}(1-X_j)$ is the number of neighbors of $i$
playing strategy $A$.

We now describe the economic model for the agents. Recall that the payoff for a $A-A$ edge is $q_A$, for a $B-B$ edge is
$q_B$ and for a $A-B$ edge is $0$. We assume that if an agent chooses $A$, his payoff is just the sum of the
payoffs obtained on each of his incident edges but if he chooses $B$, his payoff is the sum of these payoffs increased by an amount $u\geq 0$ plus a bonus of $r\geq 0$. Now the total payoff for an agent is given by
\begin{eqnarray}\nonumber
S^A_i&=&q_A \sum_{j\sim i}(1-X_j)\quad \mbox{ for strategy $A$,}\\
S^B_i&=&r+(q_B+u)\sum_{j\sim i}X_j\quad \mbox{ for strategy $B$.}\label{eq-strategies}
\end{eqnarray}


We consider that
$q_A$ and $q_B$ are fixed and correspond to the level of performance of the technologies $A$ and $B$.

By (\ref{eq-strategies}), we have $S^B_i \leq S^A_i$ iff
\begin{equation}\label{eq-payoffs}
r+(q_B+u) \sum_{j\sim i} X_j\leq q_A\sum_{j\sim i}(1-X_j)
\quad \Leftrightarrow \quad \sum_{j\sim i} X_j \leq \theta(d_i),
\end{equation}
with $\theta(d):= \frac{q_A d-r}{q_A+q_B+u}$ and $d_i$ is the degree (number of neighbours) of $i$.
We now explain the dynamics of our model for the
spread of strategy $B$ in the network as time $t$ evolves. We consider
a fixed network $G$ (not evolving in time) and let all agents play $A$
for $t<0$. At time $t=0$, some agents are forced to strategy
$B$. These agents will always play strategy $B$, hence the dynamics
described below does not apply to these initially forced agents.
We encode the initial population forced to strategy $B$ by a vector
$\chi$, where $\chi_i=1$ if agent $i$ is forced to $B$ and $\chi_i=0$
otherwise. We will assume that the vector $\chi=(\chi_i)_{i\in V}$ is a
sequence of i.i.d. Bernoulli random variables with parameter $\alpha$.

\section{Continuous-time dynamic}

We first consider the following continuous version of the contagion model. Assume that each non infected node $i$ updates its state at rate $1$ and it holds on to strategy $A$ if (\ref{eq-payoffs}) is satisfied and switches to B if $\sum_{j\sim i}X_j(t) > \theta(d_i)$. The state at time $t$ is represented by a vector $X(t)$.
Denote by $A$ the adjacency matrix of the graph $G$ and let $\lambda_1(A)$ the spectral radius of $A$, namely, its largest eigenvalue and by $d_{min}$ the minimum degree of graph $G$. In addition we will assume that the graph is connected so that $d_{min}\geq 1$ and  $\lambda_1(A)$ has multiplicity one. Therefore we have, $X_i(0)=\chi_i$, for all $i\in V$, and
$$X_i: 0 \rightarrow 1 \mbox{ \ at rate } \ind\left(\sum_j A_{ij} X_j(t) > \theta(d_i)\right) .$$

Note that $\ind\left(\sum_j A_{ij} X_j(t) > \theta(d_i)\right) \leq \frac{\sum_j A_{ij} X_j(t)}{\theta(d_i)} .$
We now consider the continuous time Markov process $Z(t)=(Z_i(t))_{i\in V}$, with $Z(0)=X(0)$, and transition rate: \\
 $$Z_i: k \rightarrow k+1 \mbox{ \ at rate \ } \frac{\sum_{j=1}^n A_{ij} Z_j(t)}{\theta(d_{min})} ,$$
standard coupling arguments yield $X(t) \leq_{st} Z(t)$ for all $t\geq 0$, where $X \leq_{st} Z$ denotes that $Z$ stochastically dominates $X$. This implies that $\sum_{i=1}^n \EE(X_i(t))\leq \sum_{i=1}^n \EE(Z_i(t))$.
Moreover, the transition rates of the process $Z$ are such that
$$\frac{d\EE[Z(t)]}{d t} = \frac {A }{\theta(d_{min})}\EE [Z(t)],$$
Hence
\begin{equation}\label{eq-coupling}
\EE[Z(t)] = e^{\frac{t}{\theta(d_{min})}A}~\EE[Z(0)].
\end{equation}
Using Cauchy-Schwartz inequality, we obtain that $\sum_{i=1}^n \EE(Z_i(t))\leq ||\EE(Z(t))||_2||1||_2$. Combining this with
(\ref{eq-coupling}), we have that
\begin{theorem}
\label{thm:continuous}
Let $\beta(t)$ be the proportion of nodes that opted for strategy $B$ by time $t$. Then
\begin{eqnarray*}
\beta(t) := \frac{\sum_{i=1}^n \EE(X_i(t))}{n} \leq \alpha e^{\frac{\lambda_1(A)}{\theta(d_{min})}t}.
\end{eqnarray*}
Moreover if the $G$ is a regular graph with degree $\Delta$, then, using the spectral decomposition of the matrix $e^{\frac{t}{\theta(\Delta)}A}$, we have that
$$\beta(t)\leq \frac{\alpha}{\Delta} e^{\frac{\Delta}{\theta(\Delta)}t}$$
\end{theorem}

The above result states that the number of nodes that have adopted $B$ increases at most exponentially in time and that the speed is given by $\frac{\lambda_1(A)}{\theta(d_{min})}$. Similar results have been found in \cite{gmt05} in the case of the Susceptible-Infected-Susceptible (SIS) epidemic.

As a matter of example for  Erd\"{o}s-R\'enyi graphs $G(n,p)$ with parameters $n$ and $p$ in the regime $\log(n)<<np$, Theorem \ref{thm:continuous} yields
$\beta(t) \leq \alpha e^{\frac{np}{\theta
(np)}t}$ with high probability.

In the next section, we describe a discrete-time version of our contagion model for which we derive more accurate results for the proportion of $B$-adopters and illustrate the coexistence of the two strategies.

\section{Discrete-time dynamic}
The state of the network at time $t$ is described by the
vector $(X_i(t))_{i\in V}$, $t\in \NN$. We have $X_i(0)=\chi_i$ and $X_i(t)\geq \chi_i$.
Then at each time step $t\geq 1$, each agent applies the best-response
update: if $S^B_i > S^A_i$ then he chooses $B$ and if not then he
chooses $A$. It is readily seen that
\begin{eqnarray}
\label{eq:rect}1-X_i(t+1)&=& (1-\chi_i) \ind\left(S^B_i(t) \leq
  S^A_i(t)\right).
\end{eqnarray}

\subsection{Diffusion process on the infinite regular tree}\label{sec:tree}
Let $T(\Delta)$ be an infinite $\Delta$-regular tree with nodes
$\o,1,2,\dots$, with a fixed root $\o$.
For a node $i$, we denote by
$\gen(i)\in \NN$
the generation of $i$, i.e. the length of the minimal path from $\o$
to $i$. Also we denote $i\to j$ if $i$ belongs to the children of
$j$, i.e. $\gen(i)=\gen(j)+1$ and $j$ is on the minimal path from
$\o$ to $i$.
For an edge $(i,j)$ with $i\to j$, we denote by $T_{i\to j}$ the
sub-tree of
$T$ with root $i$ obtained by the deletion of edge $(i,j)$ from $T$.

For a given vector $\chi$, we say that node $i\neq \o$ is
infected from $T_{i\to j}$ if the node $i$ switches to $B$ in $T_{i\to
  j} \bigcup \{(i,j) \}$ with the same vector $\chi$ for $T_{i\to j}$ and the
strategy A for $j$.
We denote by $Y_i(t)$ the corresponding indicator
function with value $1$ if $i$ is infected from $T_{i\to j}$ at time
$t$ and $0$ otherwise.
\begin{proposition}\label{lem:XY}
We have
\begin{eqnarray}
\label{eq:XY}1-X_\o(t+1) = (1-\chi_\o) \ind\left(\sum_{i\sim
    \o}Y_i(t)\leq \theta(\Delta) \right).
\end{eqnarray}
\end{proposition}

The representation (\ref{eq:XY}) is crucial to our analysis.
In fact, thanks to the tree structure, the random variables $(Y_i(t),i\sim \o)$
are independent of each other and identically distributed.
More precisely, a simple induction shows that (\ref{eq:rect}) becomes,
for $i\neq\o$:
\begin{eqnarray}
\label{eq:recttree}1-Y_i(t+1) &=& (1-\chi_i) \ind\left( \sum_{j\to i} Y_j(t) \leq \theta(\Delta)\right).
\end{eqnarray}
Note that (\ref{eq:recttree}) allows to compute all the $Y_i(t)$
recursively, starting with $Y_i(0)=\chi_i$. It is then easy to compute their distribution
from (\ref{eq:recttree}). We summarize this result in the next
proposition.

\begin{proposition}\label{prop:ht}
For $t$ fixed, the sequence $(Y_i(t), i\sim \o)$ is a sequence of
i.i.d. Bernoulli random variables with parameter $h(t)$ given by $h(0)=\alpha$ and, for $t\geq 0$,
\begin{eqnarray*}
h(t+1)=\PP(Y_i(t+1)=1) = 1 - (1-\alpha) g_{\Delta-1,\theta(\Delta)}(h(t)),
\end{eqnarray*}
where $g_{k,s}(x) = \mathbb{P}\left( Bin (k,x) \leq s \right)$. $Bin (k,x)$ corresponds to the binomial distribution with parameters $k$ and $x$.
\end{proposition}

Combining Propositions \ref{lem:XY} and \ref{prop:ht}, we obtain that
\begin{theorem}
$X_\o(t)$ is a Bernoulli random variable with parameter $\th(t)$ given
by
\begin{equation}\label{eq-tilda-h}
\PP(X_\o(t+1)=1) = \th(t+1) = 1 - (1-\alpha) g_{\Delta,\theta(\Delta)}(h(t)).
\end{equation}

Moreover let $h^*$ the smallest solution of the following fixed point equation
\begin{eqnarray}
\label{eq:fixpoint} h = 1 - (1-\alpha) g_{\Delta-1,\theta(\Delta)}(h) .
\end{eqnarray}
Suppose $0 \leq \theta(\Delta) < \Delta -2$.
There exists $\alpha_{crit}<1$ such that for all $\alpha>\alpha_{crit}$, the fixed point equation (\ref{eq:fixpoint}) has a unique solution $h^*=1$ and for all $\alpha < \alpha_{crit}$ it has three solutions $h^*<h^{**}<h^{***}=1$.
\end{theorem}

\subsection{Random regular graphs}

We now come back to the process $(X_i^{(n)}(t))_{i\in V}$ on
$G^{(n)}_\Delta$, a random $\Delta-$regular graph, satisfying (\ref{eq:rect}).
Given $d\geq 1$, let $N(i,d,G^{(n)}_\Delta)$ be the set of vertices of
$G^{(n)}_\Delta$ that are at a distance at most $d$ from $i\in G^{(n)}_\Delta$.
A depth-$d$ $\Delta$-regular tree $T(\Delta,d)$ is
the restriction of $T(\Delta)$ to nodes $i$ with $\gen(i)\leq d$.
A simple induction on $t$ shows that $X^{(n)}_i(t)$ is determined by
the $\{\chi_j, j\in N(i,t,G^{(n)}_\Delta)\}$.
Using the following convergence \cite{jlr}: for any
fixed $d\geq1$, we have as $n$ tends to infinity $N(0,d,G^{n}_{\Delta})
\stackrel{d}{\rightarrow} T(\Delta,d)$, we have
$X_0^{(n)}(t)\stackrel{d}{\rightarrow} X_\o(t)$ as $n$ tends to
infinity. Therefore the process defined on the tree in Section
\ref{sec:tree} is a good approximation of the real process. Hence,
\begin{proposition}\label{prop:limt}
For any fixed $t\geq 0$, we have
\begin{equation}
\label{eq:lim1}\lim_{n\to \infty}\EE\left[ X_i^{(n)}(t)\right]=\th(t)
\end{equation}
where $\th(t)$ is defined in (\ref{eq-tilda-h}).
\end{proposition}
Let $\beta^{(n)}(t)$ be the proportion of agents choosing $B$ at time $t$:
$\beta^{(n)}(t) = \sum_i X^{(n)}_i(t)/n$. We have as $n\to\infty$,
\begin{eqnarray}
\label{eq:mf1}\EE\left[ \beta^{(n)}(t)\right] = \EE\left[
  X_i^{(n)}(t)\right]\to\th(t).
\end{eqnarray}
The final proportion of agents choosing $B$ is $\beta^{(n)}=\lim_{t\to
\infty}\beta^{(n)}(t)$.
\begin{proposition}\label{prop:limn}
We have
\begin{eqnarray}
\label{eq:limbeta}\lim_{n\to\infty}\EE\left[ \beta^{(n)}\right]= \th,
\end{eqnarray}
in particular, for $\alpha\geq \alpha_{crit}$, we have
\begin{eqnarray}
\label{eq:limbeta1}\lim_{n\to\infty}\EE\left[ \beta^{(n)}\right]= 1.
\end{eqnarray}
\end{proposition}
The interchange of limits in $t$ and $n$ needs a proper mathematical proof. This has been done in \cite{balpit07} and
  our statement follows from their Theorem 1. For $\Delta$-regular
  graphs, bootstrap percolation is equivalent to our model.
It is noticed in \cite{balpit07} that the critical value on the
$\Delta$-regular random graph turns out to be the same as that on the
$\Delta$-tree, although the proof goes
along a quite different route.

\subsection{Extensions: random networks and linear threshold model}
In this section, we show how our approach extends to random networks and to the linear the linear threshold model (see \cite{lel} for a rigorous proof).
Let us assume that the graph $G^{(n)}$ is defined via its degree sequence $(D_i)_{i\in V}$ which is i.i.d. distributed according to $D$. Such graphs can be generated using the configuration model \cite{vanderHofstad}. Let $L_n=\sum_{k=1}^n D_k$, where $L_n/2$ is the number of edges in the graph. The underlying tree rooted at node $i$ can be described by a branching process with the offspring distribution of the root given by $D_i$. Besides the subsequent generations have offspring distribution
$$p_j^{(n)}=\sum_{k=1}^n \ind\left(D_k=j+1 \right)~\frac{D_k}{L_n}\:.$$
If the degree sequence is such that $\EE[D^2]$ is finite then, by the strong law of large numbers
$$\lim_{n\to\infty}p_j^{(n)}=\frac{(j+1)\PP(D_1=j+1)}{\EE D}\:, \quad a.s.$$

Let $\tP(j) = \frac{jP(j)}{\sum_j jP(j)}$ be the (asymptotic)
probability that an edge points to a node with degree $j$.
Then for any fixed $d$, the neighbourhood of radius $d$ about node $0$,
$N(0,d,G^{(n)})$ converges in distribution as $n$ tends to infinity to
a depth-$d$ a Galton-Watson tree with a root
which has offspring distribution $P$ and all other nodes have offspring
distribution $P^*$ given by $P^*(j-1) = \tP(j)$ for all
$j\geq 1$. Thus the associated fixed point equation is:
\begin{eqnarray}
\label{eq:fph^*}h^* = 1-(1-\alpha)\sum_j P^*(j)g_{j, \theta(j+1)}(h^*),
\end{eqnarray}
and we have $\lim_{t\to \infty} \beta(t) =\th$ given by
\begin{eqnarray}
\label{eq:th}\th = 1-(1-\alpha) \sum_j P(j)g_{j,\theta(j)}(h^*).
\end{eqnarray}
As a matter of example, for Erd\H{o}s-R\'{e}nyi graphs, the fixed point equation
associated with our model is given by (\ref{eq:fph^*}) and (\ref{eq:th}) with $P(j)=P^*(j)=e^{-\lambda}
    \frac{\lambda^j}{j!}$.
    
We consider the general threshold model \cite{kl07}. We have a non-negative random weight $W_{ij}$ on each edge, indicating the influence that $i$ exerts on $j$. We consider the symmetric case  where $W_{ij}=W_{ji}$ and we assume $W_{ij}$ are i.i.d with distribution function $W$. Each node $i$ has an arbitrary function $f_i$ defined on subsets of its neighbours set $N_i$: for any set of neighbours $X \subseteq N_i$, there is a value $f_i(X)$ between $0$ and $1$ which is monotone in the sense that if $X \subseteq Y$, then $f_i(X) \leq f_i(Y)$. This node chooses a threshold $\theta_i$ at random from $[0,1]$ and at time step $t+1$ it becomes active, it plays B, if its set of currently active neighbours $N_i^B(t)$ satisfies $f_i(N_i^B(t)) > \theta_i$.

\section{Conclusion and future work}
In this paper we presented two models of marketing wherein individuals, represented by a graph structure, receive payoffs to entice them to adopt a strategy that is different from their initial choice. To this end we initially force a small proportion of nodes to opt for the new strategy and then use an economic model that accounts for the cascading dynamic of adoption. We analyze the evolution of the proportion of agents that switch to the new strategy over time. First, the implications of our results concern marketing strategies in online social networks. More precisely, let $\alpha =\frac{\sum_i\chi_i}{n}$ be the proportion of forced
agents and let $M_1(\alpha)$ the price incurred to force the
initial agents. Typically if there is a fixed cost per agent, say $c$,
we could take $M_1(\alpha)= c\alpha$. Let $\beta(t)$ be the proportion of agents choosing $B$ at time $t$:
$\beta(t) = \frac{\sum_i X_i(t)}{n}$. We have
$\gamma(t)=\beta(t)- \alpha\geq 0$ which corresponds to the proportion of
agents choosing $B$ without being initially forced. We denote by
$M_2(\gamma(t))$ the price incurred by the rebates until time $t$.
We typically take $M_2(\gamma) = r \gamma$. Let $\delta(t)$ be the proportion of edges $B-B$ at time t.
We denote by $M_3(\delta(t))$ the price incurred by the marketing of edges until time $t$.
We typically take $M_3(\delta)= u\delta$.
Hence the total price of the marketing strategy at time $t$ is given
by $M(t)=M_1(\alpha)+M_2(\gamma(t))+M_3(\delta(t))$. One can compute the quantities $\gamma(t)$
and $\delta(t)$ in function of $\alpha,r$ and $u$. This opens the possibilities of doing an
optimal control of the marketing policy. 


Finally we remark that the marketing problem that we considered in this paper is just one application of our method.  Our approach can indeed be adapted to the analysis of the  dissemination of new versions of existing protocols, 
voting protocols through simple majority rules, i.e., $\theta(d)=\frac{d}{2}$ and distributed digital preservation systems \cite{mrgrb}.

\bibliographystyle{abbrv}
\bibliography{cont}

\end{document}